\documentclass[12pt]{iopart}

\begin{document}

\title{Rational paradigm of plasma physics}

\author{V.I.~Erofeev}
\address{Institute of Automation \&
Electrometry\\Siberian Branch of the Russian Academy of Sciences\\
1 Koptyug Prosp., 630090, Novosibirsk, RUSSIA}
\date{}

\begin{abstract}

In recent studies we have multiply envisioned the irrationality of
the traditional plasma kinetis~[1--5]. Its basic false cornerstone
was shown to be the substitution of real plasmas by plasma
probabilistic ensembles~[1--3]. We have created culture of plasma
studies with refrain from an ensemble substitution~[1-8]. We
discovered intense decay of Langmuir quanta, as opposite to
traditional deduction of their conservation~[1-3]. The quanta
decay prevents formation of Langmuir condensate in a weak Langmuir
turbulence [1, Appendix], suppresses long-wavelength plasma
modulational instability~[3], precludes Langmuir wave
collapses~[1-3]. Besides, it helped to highlight the uselessness
of hydrodynamic modelling of plasma nonlinear phenomena [2-3],  to
state an inadequateness of modelling collisionless plasma by the
Vlasov equation~[3], to state of inadequateness of ``Particle in
Cell'' method of numerical plasma simulation to the plasma nature
[3]. Finally, it was shown that existed theory undervalued notably
the intensity of induced scattering of short Langmuir waves [4]
and distorted physics of the electron distribution evolution
during this scattering [8]. With this, an extensive revision of
the plasma physics and of the theoretical physics in general are
needed.

\end{abstract}


\section{Introduction.}

Many branches of modern theoretical physics were created within
tradition of \emph{studies of statistics of physical system
ensembles.} Particularly, this can be said about \emph{all} the
current plasma physical theory. An appear of the given tradition
was \emph{conditioned} by the culture of theorizing that was
developed in physics at the end of XIX-th century: Theorists tried
then to substantiate empirical laws of thermodynamics by means of
\emph{a classical analytical mechanics.} Having absolutized notion
of a \emph{a mechanical system}\footnote[1]{This term denotes an
object possessing by finite total number of \emph{``degrees of
freedom''} and hence permitting formally the full integration of
motion equations}, they attempted to unravel the establishing of a
heat macrophysics within  a microdynamic evolution of the system.
This path was fraught with insuperable difficulties, wherefore
single finite systems were ultimately substituted by probabilistic
system ensembles\footnote[2]{As a first important system ensemble
study we can mention renown equation by
Boltzmann~\cite{Boltzmann}. His concept of \emph{distribution
function} supposes indirectly the substitution of a real mixture
of discrete gas atoms by a continuous probabilistic ensemble of
such mixtures. Boltzmann himself had no perception of this fact,
although afterwards he explicitly spoke about the desirability of
substituting system ensembles for real
systems~\cite{Boltzmann_1}}. This substitution was a basic idea
that has led Gibbs to creation of \emph{a statistical
thermodynamics (mechanics)}~\cite{Gibbs}. In view of the success
of his theory, the scientific community have rather quickly
accepted corresponding ideas as a basis  for physical theoretical
studies. They have exerted an important influence on the
development of quantum mechanics, they were successfully applied
to problem of electric conductivity of metals, to some items of
solid state physics, physics of dielectrics, physics of magnetic
materials, \emph{etc.} The highest point in this motion consisted
of developing of \emph{a nonequilibrium statistical mechanics} and
of \emph{a physical kinetics}~[12--15]. It can be ultimately said
that during the lapsed century the idea of the ensemble
substitution has been gradually recognized as one of the most
important conjectures of human genius regarding the concealed
microscopic beginnings of macrophysical phenomena in nature.
\emph{However, the latter is far not the case.}

The intention of the given our paper consists of the informing the
reader about appear of a number of plasma kinetic studies that
visualized the \emph{senselessness} of plasma ensemble
considerations and hence the \emph{irrationality} of the earlier
general tradition. (Correspondingly, the modern theoretical
physics should be subjected to an extensive revision, and
\emph{the reader is invited to this challenging and blessed
mission.})

Our motion is organized as follows. In the next section we narrate
the basis of our former arguing. In Section~3 we list results that
we developed after perceiving the irrationality of the ensemble
method and creating alternative culture of plasma studies.
Finally, Conclusions contain extra comments to the essence of
statistical mechanics and some proposals for reorganization in
theoretical physics.

\section{Senselessness of plasma ensemble studies.}

First vague suspicion of the irrationality of the ensemble method
was perceived by the given author even in years of scholarship:
\emph{The ease} was confusing with that the plasma theorists had
always spoken about the plasma ensembles and had implemented the
ensemble averaging in their calculations. Especially in studies of
plasma evolution: The Universe do not contain any evolving
probabilistic plasma ensembles. In view of this, the author
created new culture of plasma kinetic studies, one with refraining
from the plasma ensemble substitution~\cite{Yerof,Yerof_1}.
Applying it to problems of plasma turbulence theory, the author
has found first important illustration of the inconsistency of the
ensemble substitution: The physics of collisionless wave
dissipation. It was a common belief that quanta of Langmuir wave
field can decay in a collisionless plasma only due to the Landau
damping~\cite{Landau}, whereas new calculations visualized intense
decay of Langmuir wave quanta due to \emph{the nonresonant
electron diffusion in momentums} (it is called also \emph{the
stochastic electron acceleration}).

Trials to publish corresponding calculations have met an immense
scepsis of many plasma theorists. Even the pointing at a
respective contradiction within the very body of traditional
plasma theory did not had an adequate effect\footnote[3]{We mean
the following. Rigorous recognized calculations of former
ensemble-oriented weak plasma turbulence theory comprised not the
conclusion of the Langmuir quanta conservation only but equally
the deduction of their intense decay, see details in
Refs.~\cite{Yerof4,Yerof6}}. This forced the author to seek for
experimental confirmation of his conclusions. This confirmation
was found in existing reports on beam-plasma experiments with
``strong'' Langmuir turbulence. Following widespread belief there
should have been observed Langmuir wave collapses, whereas the
anew deduced Langmuir quanta decay precluded their initiation. The
author have thoroughly analyzed three independent series of
reported ``experimental collapse observations'' and have proven
that they evidenced not for the collapse but just against
it~[1--3]. This ultimately helped to affirm the key idea:
\emph{Laws of physical evolution of statistics of a plasma
ensemble depend on the ensemble content.} Correspondingly, one
should not hope on approaching the objective picture of the plasma
physical phenomena when he substitutes a real plasma by an
imaginary plasma ensemble.

We have discussed in this section the most important of currently
developed illustrations of inappropriateness of the ensemble
method in plasma studies: The physics of Langmuir turbulence
dissipation constituted one of the cornerstones in the former
plasma turbulence theory. In the next section we supplement this
illustration with other our results that have also visualized the
\emph{uselessness} of plasma ensemble considerations. Besides, we
present there an outline of some more general consequences of our
discoveries for the current practice of plasma theoretical
studies.

\section{List of achievements in studies of ``single'' plasmas.}

Development of a real physical picture of the collisionless
Langmuir turbulence dissipation that we mentioned above has
permitted us to discover three important consequences.  First, the
dissipation impedes the formation of the Langmuir wave condensate
(see Appendix in Ref.~\cite{Yerof4})\footnote[4]{The reader is
reminded that the Langmuir condensate was believed to form due to
the wave scattering induced by plasma particles: waves should have
been ultimately transferred to the long-wavelength region of the
turbulence spectrum. The puzzle of dissipation of the condensate
energy has motivated large body of former theoretical plasma
turbulence studies}. Second, it suppresses the Zakharov's  short
-wavelength plasma modulational instability~\cite{Yerof6}. This
instability was originally rendered as an initial stage of
Langmuir wave collapse~\cite{Zakhar}, hence its absence just
highlights the impossibility of the collapse
initiation\footnote[5]{Note that the ``collapse science'' was a
rather opulent branch of former plasma turbulence theory, and its
contributors developed pictures of collapse scenarios even without
the above short -wavelength modulational instability. For this
reason we have supplemented the earlier study by thorough analyzes
of diverse items of that science and ultimately affirmed the total
inconsistency of the collapse concept~\cite{Yerof10}. Some
conclusions of the corresponding analyzes we mention in this paper
further on}. Third, the real turbulence collisionless dissipation
prevents the development of Vedenov -Rudakov's long-wavelength
plasma modulational instability~\cite{Yerof10}\footnote[6]{The
reader is reminded that corresponding paper by Vedenov and
Rudakov, see Ref.~\cite{Veden}, was one of the most stimulating
plasma theory publications}. It is worth emphasizing that the
concepts of the Langmuir condensate, of the Vedenov -Rudakov's
plasma modulational instability and of the Zakharov's
short-wavelength plasma modulational instability constituted the
kern of the former Langmuir turbulence theory. Their failures
illustrate rather well the senselessness of plasma ensemble
considerations.

The irrationality of the ensemble method was highlighted also by
our results on wave scattering induced by plasma
electrons~\cite{Yerof5,Yerof11}. On the one side, we have shown
that formerly accepted Tsytovich's picture of the phenomenon (see
Ref.~\cite{Tsyt_4}) undervalued substantially the intensity of
induced scattering of short Langmuir waves~\cite{Yerof5}. In full
accord with the common sense, the plaster of ensemble averaging has
veiled the genuine fresco of the phenomenon. On the other side, it was
found that corresponding vision of the process does not simply
veil substantially the genuine evolution of electron distribution
during the scattering, but entirely distorts its
picture~\cite{Yerof11}.

In such a way, our plasma kinetic considerations have multiply
envisioned \emph{the senselessness} of plasma ensemble studies.
Note that this means an \emph{inappropriateness} of the recipes of
\emph{nonequilibrium statistical mechanics} for the plasma
studies. The latter theoretical discipline is \emph{absolutely
useless} for revealing objective laws of plasma physical behavior.

We reinforce that the nonequilibrium statistical mechanics had
exerted a {\em key\/} influence on formation of the most basic of
the plasma physical theoretical notions: It underlies the plasma
kinetics after Bogoliubov~\cite{Bog}
--- Born --- Green~\cite{BornGreen} --- Kirkwood~\cite{Kirk} ---
Yvon~\cite{Yvon} (that is collectively known as the BBGKY
kinetics), it is hidden behind any hydrodynamic plasma modelling,
\emph{etc.} Further, it was always rendered as a basis for many
other physical theories apart from the plasma theory. Plasma
phenomena are surely not the exclusion from the common rule, hence
our observations unravelled \emph{an absolute truth} of \emph{a
scientific inconsistency} of the nonequilibrium statistical
mechanics: Its scenarios have generally no correlation with real
pictures of physical system evolutions. Note that the current gas
kinetic theory seems to contain an independent item that might
have been shown to confirm this conclusion. It is the above
mentioned Boltzmann equation: We suspect that real gas kinetic
evolution deviate from scenarios given by the equation. On the one
side, some serious problems with the Boltzmann equation were
enlightened even in 1876: consistent critical comments regarding
the equation are known after Loshmidt~\cite{Loshmidt},
Zermelo~\cite{Zermelo} and Ehrenfests~\cite{Ehrenfests}. On the
other side, it was recently discovered  by
Alekseev~\cite{Alekseev} that assumption  of finite (nonzero)
duration of a pair atom collision results in corrections to the
Boltzmann equations that notably modify the usual terms of the
equation.

In the given paper we cannot discuss details of our above
mentioned studies: The reader is addressed to original
papers~[1--8] for  more expanded description of our ideas,
approaches and achievements. We complete this section by returning
to plasma physics and  mentioning of some more of our deductions
relevant to the practice of plasma theoretical studies.

Consider the BBGKY kinetics. Outline of generic defects of the
given theory the reader can find in Conclusions of
paper~~\cite{Yerof7}. The main of them is that the conceptions of
this kinetics acquired some sense of constructive theory only
within the practice of the plasma ensemble considerations.
Correspondingly, its recipes cannot help one in the exploring  the
objective physics of plasma phenomena.

Consider the practice of hydrodynamic plasma studies.  In any
hydrodynamic plasma description the plasma ensemble is usually not
directly referred to but factually implied, since hydrodynamics do
not distinguishes any real plasma from many others with slightly
varied positions and momentums of individual plasma particles and
hence from many of plasma ensembles. With this, the currently
popular hydrodynamic modelling of nonlinear plasma phenomena has
no much sense for explaining plasma physical realities: The degree
of correspondence of any hydrodynamic model to plasma realities is
always doubtful. Both theoretical predictions of new physical
effects and trials to explain experimental observations, with
exclusive appeal to a certain nonlinear hydrodynamic plasma model,
are scientifically unsound. These positions are rather well
illustrated by failure of Zakharov's plasma hydrodynamics with its
prediction of the Langmuir wave collapse, and also by failure of
Vedenov -Rudakov's hydrodynamic  plasma model with its prediction
of the long-wavelength plasma modulational instability.

\

{\bf Finally}, exploration of problems of ``collapse science'' has
inspired the author on

\begin{itemize}
  \item [A)] showing the impossibility of wave energy accumulation in plasma density
  cavities via nucleation~\cite{Yerof10}\footnote[7]{For the first time
  the misconception of \emph{nucleation} was
  advanced by Doolen~\cite{Doolen}, on a basis of one-dimensional simulation of
  Zakharov's (nonlinear) plasma hydrodynamics~\cite{Zakhar}},

  \item [B)] stating of inadequateness of modelling the collisionless plasma by the
Vlasov equation instead of the Klimontovich -Dupree's
one~\cite{Yerof10}\footnote[8]{Correspondingly, we developed more
adequate understanding of applicability area of the Vlasov
equation in plasma considerations. One of the equation
applications that is conceptually  acceptable for the nonlinear
plasma studies the reader can reveal in intermediate calculations
of Ref.~\cite{Yerof7}},

  \item [C)] stating of inadequateness of the ``Particle in Cell''
  method of numerical plasma simulation to the plasma
  nature~\cite{Yerof10}.

\end{itemize}

\section{Conclusions}

Following deductions reported in preceding sections, the practice
of plasma theoretical studies (and hence also the practice of
interpreting of experimental plasma observations) requires an
extensive revision. The reader is invited to explore our
approaches from original papers~[1--8] and then to join us in the
activity of reestablishing the plasma physics.

As we are persistently reiterating, our findings are extremely
important not for the plasma physics only but for all the
theoretical physics. Having visualized the irrationality of the
ensemble method, we restored common sense that was always ignored
during its erection. Note that in Conclusions of
Ref.~\cite{Yerof10} we have rather thoroughly commented the
history of ensemble studies and their trend in physics.
Particularly, we have commented there the reasons for the success
of the gibbsian statistical mechanics. This time we will not
repeat respective comments: We supplement them with a number of
natural final conclusions.

First, the reader should understand that the statistical mechanics
is a mere artificial theory that \emph{emulates} only, although
rather successfully, the laws of thermodynamics. The
appropriateness of the gibbsian ensemble substitution can be
rationally justified only for homogeneous thermodynamically
equilibrium spatially extensive physical medium whereat the
mutually remote pieces might be rendered as ``independent ensemble
realizations''\footnote[9]{Note also that for thermodynamic
phenomena the idea of a single system is rather inadequate
abstraction: One can theorize at best about subsystem that is a
part of a larger subsystem and that constantly exercises an
influence from the side of remaining part of the latter.
Naturally, corresponding detailed influence of one on another
never can be deduced, but even this do not justifies the
substitution of a certain subsystem by a probabilistic ensemble of
subsystems. Besides, even the very notion ``subsystem'' is
speculative a bit. Really, some portions of atoms (or molecules,
or electrons, \emph{etc.}) do inevitably leave the subsystem each
moment, others enter it from the surroundings, whereas the
beginnings of statistical mechanics do not yield this permanent
renewal of the subsystem}. With this, it is not only that the
nonequilibrium statistical mechanics should be ejected off the
building of modern physics, but also the basic positions of
\emph{the equilibrium statistical mechanics} should be thoroughly
revised, at least in the sense of their interpretations.
Particularly, we can say that existing pretension of the
statistical mechanics on revealing of the hidden essence of
thermodynamical entropy is far not sound. Really, the entropy of
statistical mechanics is defined as a characteristic of a system
ensemble, and its non-lessening during the ensemble evolution is
postulated. But laws of the system ensemble evolution depend on
the ensemble content and cannot provide one with the objective
picture of the certain system physical evolution. (For instance,
of the thermalization of initially nonequilibrium gas in a fixed
volume.) Correspondingly, the growth of the statistical mechanical
entropy has generally no correlation with the corresponding growth
of the thermodynamical entropy. Moreover, while the latter is well
measurable and hence is always definite, the former depend on the
content of the artificially introduced gas ensemble. In short, the
sole virtue of the statistical mechanical entropy is that its
changes in \emph{a reversible thermodynamic gas process} coincides
with the respective changes of the thermodynamical entropy,
provided one discusses either the gibbsian canonical, the big
canonical or the microcanonical gas ensembles. It was just due to
this property that creators of the statistical mechanics
absolutized the corresponding entropy.

Second, by the earlier set of deductions regarding the plasma
physical theory we have visualized a necessity of creation \emph{a
physics of discrete media} that should supersede the existing
\emph{physics of continuous media.} Following the traditional
culture of plasma physics, researches are suggested to render
plasmas  as continuous media: Key notions of plasma theory were
originally created within the physics of continuous media and then
adjusted to peculiar problems of plasma studies. Particularly,
common notion of \emph{a wave},  plasma natural oscillation,
suggests attitude to a plasma as a continuous medium. This
attitude was embodied by the concept of \emph{a Vlasov plasma:}
Vlasov equation comprises imaginary electron and ion liquids that
fill without gaps the 6-dimensional phase space. Having imagined a
Vlasov plasma, one substitutes indirectly the real mixture of
individual charged particles, known as \emph{a ``Klimontovich
-Dupree'' plasma,} by an ensemble of such mixtures. It was an
understanding of drastic distinction in physics of these two
objects that have ultimately helped us in developing of earlier
impressive paradigmatic deductions. We suppose that the visualized
substantial  distinction between the physics of the Vlasov plasma
and that of the Klimontovich -Dupree one exemplifies the general
distinction between the objective physical pictures of discrete
medium (to be developed) and their images from the physics of
continuous media. Besides, we propose to create the methodology of
earlier physics of discrete media by an analogy with approaches of
our plasma kinetics.

Finally, our discoveries have highlighted the necessity of
revision of the  \emph{physical kinetics.} This science was always
assumed to include the traditional plasma kinetics as a daughter
branch, whereas we have multiply demonstrated the falseness of the
basis of the latter.

\

\end{document}